\documentclass[12pt,a4paper]{article}

\usepackage[nosort]{cite}
\usepackage[hyperref,bulletsep]{collect}
\usepackage{amsmath,amssymb,graphicx}
\usepackage{pst-all}
\usepackage{bbm}

\makeatletter \@addtoreset{equation}{section} \makeatother

\makeatletter
\let\old@startsection=\@startsection
\let\oldl@section=\l@section
\renewcommand{\@startsection}[6]{\old@startsection{#1}{#2}{#3}{#4}{#5}{#6\mathversion{bold}}}
\renewcommand{\l@section}[2]{\oldl@section{\mathversion{bold}#1}{#2}}
\makeatother

\makeatletter
\let\old@makecaption=\@makecaption
\def\@makecaption{\small\old@makecaption}
\makeatother

\renewcommand{\leq}{\leqslant}

\newcommand{\dder}[1]{\overleftrightarrow{\partial_{#1}} }

\begin{document}

\thispagestyle{empty}
\begin{flushright}\footnotesize
\texttt{ITEP-TH-11/09}\\
\texttt{LPTENS-09/05}\\
\texttt{UUITP-08/09} \vspace{0.8cm}
\end{flushright}

\renewcommand{\thefootnote}{\fnsymbol{footnote}}
\setcounter{footnote}{0}

\begin{center}
{\Large\textbf{\mathversion{bold} Worldsheet spectrum
\\  in $AdS_4/CFT_3$ correspondence
}\par}

\vspace{1.5cm}

\textrm{K.~Zarembo\footnote{Also at ITEP, Moscow, Russia}}
\vspace{8mm}

\textit{CNRS -- Laboratoire de Physique Th\'eorique,
\'Ecole Normale Sup\'erieure\\
24 rue Lhomond, 75231 Paris, France }\\
\texttt{Konstantin.Zarembo@lpt.ens.fr} \\ \vspace{3mm} and
\vspace{3mm}

\textit{Department of Physics and Astronomy, Uppsala University\\
SE-751 08 Uppsala, Sweden}\\
\vspace{3mm}


\par\vspace{1cm}

\textbf{Abstract} \vspace{5mm}

\begin{minipage}{14cm}
The $AdS_4/CFT_3$ duality is a new example of an integrable and
exactly solvable AdS/CFT system. There is, however, a puzzling
mismatch between the number of degrees of freedom used in the exact
solution ($4_B+4_F$ scattering states) and $8_B+8_F$ transverse
oscillation modes of critical superstring theory. We offer a
resolution of this puzzle by arguing that half of the string modes
dissolve in the continuum of two-particle states once $\alpha '$
corrections are taken into account. We also check that the
conjectured exact S-matrix of $AdS_4/CFT_3$ \cite{Ahn:2008aa} agrees
with the tree-level worldsheet calculation.
\end{minipage}

\end{center}

\vspace{0.5cm}


\newpage
\setcounter{page}{1}
\renewcommand{\thefootnote}{\arabic{footnote}}
\setcounter{footnote}{0}

\section{Introduction}

A new example of the $AdS_4/CFT_3$ duality proposed in
\cite{Aharony:2008ug} and further developed in \cite{Aharony:2008gk}
establishes an equivalence of the superconformal Chern-Simons-matter
theory and type IIA string theory on $AdS_4\times CP^3$
\cite{Arutyunov:2008if,Stefanski:2008ik,Gomis:2008jt}. This AdS/CFT
system turns out to be integrable
\cite{Minahan:2008hf,Arutyunov:2008if,Stefanski:2008ik,Gromov:2008bz}
and exactly solvable \cite{Gromov:2008qe,Ahn:2008aa,Gromov:2009tv}
in the large-$N$ (free string) limit. However, certain features of
the available exact solution are rather puzzling and call for an
explanation. The purpose of this paper is to address these puzzles
and to clarify how the building blocks of the exact solution fit
together with the perturbative spectrum of the string sigma-model.

Let us begin by shortly describing the exact solution of the planar
$AdS_4/CFT_3$ system \cite{Gromov:2009tv} from the worldsheet
perspective, which is most appropriate for our purposes. An
alternative but equivalent interpretation can be given in terms of
an integrable spin chain that arises in the $\mathcal{N}=6$
Chern-Simons theory
\cite{Minahan:2008hf,Gaiotto:2008cg,Bak:2008cp,Bak:2008vd,Zwiebel:2009vb,Minahan:2009te}.
The solution of \cite{Gromov:2009tv} describes the exact spectrum of
the string sigma-model in the light-cone gauge, which is a massive
two-dimensional integrable quantum field theory. The main building
blocks of the exact solution are the spectrum of elementary
excitations and their exact S-matrix \cite{Ahn:2008aa}. The
S-matrix, defined on an infinitely long string, can be diagonalized
by the Bethe ansatz equations \cite{Gromov:2008qe}. The periodic
boundary conditions of the closed string are then imposed with the
help of the  Y-system and thermodynamic Bethe ansatz
\cite{Gromov:2009tv}.

The global symmetries, along with integrability, are of central
importance for this construction. The global symmetry group of the
string sigma-model on $AdS_4\times CP^3$ is $OSp(6|4)$, but in the
light-cone gauge only the $PSU(2|2)\times U(1)$ subgroup is linearly
realized. The worldsheet spectrum consists of massive particles,
which transform in the $(2|2)_{+1}$ and $(2|2)_{-1}$ representations
of $PSU(2|2)\times U(1)$ and have the dispersion relation
\cite{Gaiotto:2008cg}:
\begin{equation}\label{light}
 \varepsilon (p)=\sqrt{\frac{1}{4}+4h^2(\lambda )\sin^2\frac{p}{\sqrt{2\lambda }}}
 \,\,\stackrel{\lambda \rightarrow \infty
 }{\longrightarrow }\,\,\sqrt{p^2+\frac{1}{4}}\,.
\end{equation}
Here $\lambda $ is the 't~Hooft coupling of the Chern-Simons theory.
The string tension is given by\footnote{In $AdS_4\times CP^3$, the
string tension is apparently renormalized \cite{Bergman:2009zh}.
$\sqrt{\lambda /2}$ is the large-$\lambda $ asymptotics.}
$T=\sqrt{\lambda /2}$ \cite{Aharony:2008ug}. $h(\lambda )$ is an
interpolating function \cite{Gaiotto:2008cg} which behaves as
$h(\lambda )\simeq\sqrt{\lambda /2}$ at strong coupling. At finite
$p$ and $\varepsilon $ the elementary excitations are transverse
fluctuations of the string moving on the light-like geodesic (the
BMN modes)
\cite{Nishioka:2008gz,Gaiotto:2008cg,Arutyunov:2008if,Grignani:2008is}.
If one pumps momentum $p\sim \sqrt{\lambda }$ in a BMN mode, it
blows up into a soliton (the giant magnon)
\cite{Gaiotto:2008cg,Grignani:2008is,Grignani:2008te,Lee:2008ui,Shenderovich:2008bs,Ahn:2008hj,Ryang:2008rc,Bombardelli:2008qd,Lukowski:2008eq,Ahn:2008wd,Abbott:2008qd,Kalousios:2009mp,Suzuki:2009sc}.
The factorizable scattering matrix of the solitons was constructed
in \cite{Ahn:2008aa} and is built from the Beisert's
$PSU(2|2)$-invariant S-matrix \cite{Beisert:2005tm,Beisert:2006qh}
and the BES/BHL dressing phase \cite{Beisert:2006ez,Beisert:2006ib}.

This picture has two puzzling features. Firstly, the
soliton-antisoliton scattering according to \cite{Ahn:2008aa} is
completely reflectionless. This does not contradict any basic
principles, but does not follow from the known symmetries either. In
fact, different factorizable S-matrices with the same symmetries do
exist, and have a non-vanishing reflection amplitude
\cite{Ahn:2008tv}. These S-matrices, however, disagree with explicit
two-loop computations in the Chern-Simons theory \cite{Ahn:2009zg}.
Secondly, the number of degrees of freedom in the scattering theory
and on the worldsheet do not match. The scattering theory of
\cite{Ahn:2008aa} describes $(2|2)+(2|2)=(4|4)$ worldsheet degrees
of freedom, while critical superstring should have $(8|8)$. Indeed,
fixing the light-cone gauge in the sigma-model (or, alternatively,
expanding near a non-trivial spinning string solution
\cite{McLoughlin:2008ms,Alday:2008ut,Krishnan:2008zs,Gromov:2008fy,McLoughlin:2008he}),
one finds sixteen massive oscillation modes. Those include, in
addition to the light modes (\ref{light}), $(4|4)$ heavy modes with
the dispersion \cite{Nishioka:2008gz,Gaiotto:2008cg,Grignani:2008is}
\begin{equation}\label{}
 E(p)=\sqrt{p^2+1}\,.
\end{equation}
An interpretation of these modes is rather unclear. They do not show
up in the scattering theory. Does this mean that the S-matrix is
incomplete and has to be extended to a larger set of elementary
excitations? The classical, large-$\lambda $ limit of the Bethe
equations \cite{Gromov:2008bz} suggests that this is not the case.
As shown in \cite{Gromov:2008bz}, all sixteen BMN modes can be
identified as solutions to the classical string Bethe equations.
However, the solutions associated with the light and heavy modes are
qualitatively different. While the light modes are associated with
solitary Bethe roots, the heavy modes are described by
\textit{stacks} \cite{Gromov:2008bz}. The stacks
\cite{Beisert:2005di,Gromov:2007ky} are compounds of two
momentum-carrying Bethe roots, which are not bound states of the
corresponding particles. Stacks can be identified only at extreme
values of some small parameter\footnote{See an example in sec.~2 of
\cite{Gromov:2007ky}.}, in this case the inverse string tension
$2/\sqrt{2\lambda }$. At finite $\lambda $, a stack is
indistinguishable from a generic two-particle state.

We suggest the following resolution of the apparent mismatch in the
number of degrees of freedom. In order to understand if the heavy
modes exist as elementary excitations, we need to see whether they
appear as poles in Green's functions or not. Let us consider the
propagator of a heavy mode $G(\omega ,p)$. At strictly infinite
coupling it has a pole at $\omega =E(p)$. We are going to argue,
however, that $\alpha '$ corrections cannot be neglected even if the
coupling $1/\sqrt{2\lambda }$ is very small. The reason is that the
pole lies exactly at the threshold of pair production of two light
modes. The polarization operator thus has a cut with the branch
point also at $\omega =E(p)$: $\Pi(\omega ,p)\sim i(\omega-E(p)
)^\nu $. For the $\Phi \phi^2$-type interaction in two dimensions,
we would expect $\nu =-1/2$, but as we shall see later the
derivatives in the heavy-light-light interaction vertex soften the
threshold singularity to $\nu =+1/2$. The one-loop corrected Green's
function thus behaves near the threshold as
\begin{equation}\label{threshprop}
 G(\omega ,p)\sim \frac{iZ(p)}{\omega -E(p) +\frac{C(p)}{\sqrt{2\lambda }}\,\sqrt{
 E(p)-\omega }}\,.
\end{equation}
Sufficiently close to $E(p)$, for $|\omega -E(p)|\sim 1/\lambda $,
the second term in the denominator is as important as the first one
and can change the analytic properties of $G(\omega ,p)$. Whether
the pole in $G(\omega ,p)$ survives or not depends on the sign of
$C$. If $C$ is positive, the pole shifts below the threshold:
$E_{\rm 1-loop} =E-C^2/2\lambda $. If $C$ is negative, the pole
disappears. The propagator then has only a cut on the physical sheet
which means that the heavy mode dissolves in the continuum and does
not exist as a physical excitation at finite $\lambda $.

We will calculate the polarization operator for one of the heavy
modes (the others should be related by supersymmetry) using the
large-$\lambda $ (near-BMN) expansion of the sigma-model in
$AdS_4\times CP^3$ \cite{Astolfi:2008ji,Sundin:2008vt}. In
\cite{Astolfi:2008ji,Sundin:2008vt}, the near-BMN expansion was
utilized to compute energy shifts of the BMN oscillator spectrum. In
these calculations the heavy modes were integrated out. The
agreement of the near-BMN calculations with the Bethe-ansatz
predictions \cite{Astolfi:2008ji,Sundin:2008vt} implies that a
contribution of the heavy modes is already contained in the S-matrix
of \cite{Ahn:2008aa}. We will compute the worldsheet S-matrix
explicitly and in particular will verify that the reflection
amplitude in the particle-antiparticle scattering cancels out to the
leading order is the $1/\sqrt{\lambda }$ expansion.

\section{The sigma-model}

We will start with the supercoset formulation of the worldsheet
sigma-model on $AdS_4\times {C}P^3$
\cite{Arutyunov:2008if,Stefanski:2008ik,Uvarov:2008yi}, which arises
after partially fixing kappa-symmetry gauge in the full
Green-Schwarz action \cite{Gomis:2008jt}. The construction of the
supercoset $OSp(6|4)/SO(3,1)\times U(3)$ sigma-model is based on the
$\mathbbm{Z}_4$ decomposition of the $osp(6|4)$ superalgebra:
\begin{equation}\label{}
 osp(6|4)=h_0\oplus h_1\oplus h_2\oplus h_3.
\end{equation}
Since the superalgebra admits a $\mathbb{Z}_4$ automorphism
\cite{Arutyunov:2008if,Stefanski:2008ik}, this decomposition is
consistent with the (anti)commutation relations: $[h_i,h_j\}\subset
h_{(i+j)\!\!\mod\! 4}$. The $h_0$ subalgebra is the denominator of
the coset: $h_0=so(3,1)\oplus u(3)$, $h_0\oplus h_2$ is the bosonic
subalgebra $so(3,2)\oplus su(4)$, and $h_1$, $h_3$ contain all the
odd generators. The worldsheet embedding coordinates are
parameterized by a coset representative $g(x)\in OSp(6|4)$, defined
up to gauge transformations $g(x)\rightarrow g(x)h(x)$ with $h(x)\in
SO(3,1)\times U(3)$. The global $OSp(6|4)$ acts from the left:
$g(x)\rightarrow g'g(x)$. The sigma-model action can be expressed in
terms of the $OSp(6|4)$-invariant current
\begin{equation}\label{current}
 P_\mu =g^{-1}\partial _\mu g=P_{0\,\mu }
 +P_{1\,\mu }+P_{2\,\mu }+P_{3\,\mu }.
\end{equation}
The $h_0$ component of the $\mathbbm{Z}_4$ decomposition of $P_\mu $
transforms as a gauge connection: $P_{0\,\mu} \rightarrow
h^{-1}P_{0\,\mu} h+h^{-1}\partial _\mu h$. The other three
components transform homogeneously: $P_{1,2,3\,\mu} \rightarrow
h^{-1}P_{1,2,3\,\mu} h$. The Lagrangian of the sigma-model
is\footnote{We use the $(+-)$ conventions for the worldsheet metric;
the epsilon tensor is defined such that $\epsilon ^{01}=1$.}
\begin{equation}\label{Lmaster}
 \mathcal{L}=\frac{\sqrt{2\lambda }}{2}\,\mathop{\mathrm{Str}}\left(
 \sqrt{-h}h^{\mu \nu }P_{2\,\mu}P_{2\,\nu} +\epsilon ^{\mu \nu }
 P_{1\,\mu }P_{3\,\nu}
 \right),
\end{equation}
where $\mathop{\mathrm{Str}}(\cdot\,\,\cdot)$ is the invariant
bilinear form on $osp(6|4)$.

The light-cone gauge breaks $OSp(6|4)$ down to $PSU(2|2)\times
U(1)$, of which only the bosonic subgroup $SU(2)\times SU(2)\times
U(1)$ will be manifest. This suggests the following choice of basis
in $osp(6|4)$:
\begin{eqnarray}\label{basis}
 h_0&=&\{R^a_b,R^a,R_a,R;K_i,T_i\}\nonumber \\
 h_2&=&\{B^a,B_a,J,M;L_i,D\}\nonumber \\
 h_1&=&\{Q^a_\alpha ,\bar{Q}_{a\alpha },Q_\alpha ,\bar{Q}_\alpha \}\nonumber \\
 h_3&=&\{S_{a\alpha },\bar{S}^a_\alpha ,S_\alpha ,\bar{S}_\alpha \},
\end{eqnarray}
where $a,b=1,2$ and $\alpha =1,2$ are the indices of the unbroken
$SU(2)\times SU(2)$. The commutation relations of $osp(6|4)$ are
listed in appendix~\ref{osp64}, along with the components of the
invariant bilinear form. The light-cone coordinates in the coset are
conjugate to $D\pm iJ/2$. The unbroken subalgebra $su(2)\oplus
su(2)\oplus u(1)$ is the set of generators that commute with both
$J$ and $D$: $\{R^a_b-\delta ^a_bR^c_c/2,R;T_i\}$. In the limit of
an infinitely long string, which we will take later, the
supercharges that commute with the light-cone Hamiltonian $D-iJ/2$
are also conserved. Those are $Q^a_\alpha -i\varepsilon
^{ab}\bar{Q}_{b\alpha }$ and $\bar{S}^a_\alpha +i\varepsilon
^{ab}S_{b\alpha }$. Together with the conserved bosonic generators
they form a $psu(2|2)$ subalgebra.

In order to fix the light-cone gauge, it is convenient to use the
following parameterization of the coset representative:
\begin{equation}\label{}
 g=\,{\rm e}\,^{\Xi }\,{\rm e}\,^{\mathbbm{X}},
\end{equation}
where
\begin{equation}\label{}
 \Xi =tD+\frac{i}{2}\,\varphi J
\end{equation}
and
\begin{eqnarray}\label{cosetX}
 \mathbbm{X}&=&\frac{1}{\sqrt{2}}\,X^aB_a+\frac{1}{\sqrt{2}}\,\bar{X}_aB^a
 +\frac{1}{2}\,ZM+Y^iL_i+\frac{1}{2}\,\varepsilon ^{\alpha \beta }\varepsilon
 _{ab}\chi ^a_{R\,\alpha }Q^b_\beta
 +\frac{i}{2}\,\varepsilon ^{\alpha \beta }\chi ^a_{R\,\alpha
 }\bar{Q}_{a\beta }
 \nonumber \\ &&
 -i\bar{\psi }^\alpha _LQ_\alpha
 +i\varepsilon ^{\alpha \beta }\psi _{R\,\alpha }\bar{Q}_\beta
 -\frac{1}{2}\,\varepsilon ^{\alpha \beta }\chi ^a_{L\,\alpha }S_{a\beta }
 -\frac{i}{2}\,\varepsilon ^{\alpha \beta }\varepsilon _{ab}\chi
 ^a_{L\,\alpha }\bar{S}^b_\beta -\varepsilon ^{\alpha \beta }\psi
 _{L\,\alpha }S_\beta +\bar{\psi }^\alpha _{R}\bar{S}_\alpha .
\end{eqnarray}
By choosing the same coefficients in front of $Q^a_\alpha $,
$\bar{Q}_{a\alpha }$ and $S_{a\alpha }$, $\bar{S}^a_\alpha $, we
eliminate eight fermionic degrees of freedom and thus fix the
residual kappa-symmetry \cite{Arutyunov:2008if} of the coset
sigma-model.

The light-cone gauge treats $t$, $\varphi $ and the rest of the
world-sheet coordinates asymmetrically. The centre of mass of the
string moves along the light-like geodesic $\varphi =t$, and we need
to keep the dependence on $t$ and $\varphi $ in the Lagrangian
exactly. The string oscillations in the transverse directions, which
are parameterized by $\mathbbm{X}$, are small and can be treated
perturbatively. The current (\ref{current}) expands as
\begin{equation}\label{}
 P_\mu =\partial _\mu \Xi
 +\frac{1-\,{\rm
 e}\,^{-\mathop{\mathrm{ad}}\mathbbm{X}}}{\mathop{\mathrm{ad}}\mathbbm{X}}\,
 \mathcal{D}_\mu \mathbbm{X}
 =\partial _\mu \Xi +\mathcal{D}_\mu \mathbbm{X}
 -\frac{1}{2}[\mathbbm{X},\mathcal{D}_\mu \mathbbm{X}]
 +\frac{1}{6}[\mathbbm{X},[\mathbbm{X},\mathcal{D}_\mu \mathbbm{X}]]
 +\ldots .
\end{equation}
where the long derivative is defined by
\begin{equation}\label{}
 \mathcal{D}_\mu \mathbbm{X}=\partial _\mu \mathbbm{X}+[\partial _\mu \Xi
 ,\mathbbm{X}].
\end{equation}
Plugging this expansion into the Lagrangian (\ref{Lmaster}) one can
get the sigma-model action to any desired order in $\mathbbm{X}$,
although explicit expressions quickly get complicated.

To the two lowest orders, and omitting the overall factors of
$\sqrt{\lambda/2 }$ and $\sqrt{-h}$,
\begin{equation}\label{L0}
 \mathcal{L}^{(0)}=\frac{1}{2}\left(\partial _\mu \varphi
 \right)^2-\frac{1}{2}\left(\partial _\mu t\right)^2,
\end{equation}
and
\begin{eqnarray}\label{L2}
 \mathcal{L}^{(2)}&=&\partial _\mu \bar{X}_a\partial ^\mu X^a
 +\frac{1}{2}\left(\partial _\mu Z\right)^2
 +\frac{1}{2}\left(\partial _\mu Y_i\right)^2
 \nonumber \\ &&
 +i\bar{\psi }^\alpha \gamma ^\mu\dder{\nu} \psi _\alpha\Pi_\mu
 ^{\nu \lambda }\partial _\lambda t
 +\frac{i}{4}\,\bar{\chi }^\alpha _a\gamma ^\mu \dder{\nu}
 \chi ^a_\alpha \Pi_\mu ^{\nu \lambda }\partial _\lambda \left(t+\varphi \right)
 \nonumber \\ &&
 -\left(\frac{1}{2}\,Y^2_i+\frac{1}{2}\,\bar{\psi }^\alpha \psi _\alpha
 +\frac{1}{8}\,\bar{\chi }^\alpha _a\chi ^a_\alpha \right)\left(\partial _\mu
 t\right)^2
 -\left(\frac{1}{2}\,Z^2+\frac{1}{4}\,\bar{X}_aX^a
 +\frac{1}{8}\,\bar{\chi }^\alpha _a\chi _\alpha ^a\right)\left(\partial _\mu \varphi
 \right)^2
 \nonumber \\ &&
 -\frac{1}{4}\,\bar{\chi }^\alpha _a\chi _\alpha ^a\partial _\mu t\partial
 ^\mu \varphi
 -\frac{1}{8}\,\bar{\chi }^\alpha _a\chi _\alpha ^a\epsilon ^{\mu \nu }\partial
 _\mu t\partial _\nu \varphi ,
\end{eqnarray}
where the fermions have been combined into two-dimensional Dirac
($\psi _\alpha $) and Majorana ($\chi _\alpha ^a$) spinors:
\begin{eqnarray}\label{}
 \psi _\alpha =
 \begin{pmatrix}
   \psi _{L\,\alpha }   \\
   \psi _{R\,\alpha }  \\
 \end{pmatrix},&\qquad&
 \bar{\psi }^\alpha =\left(\bar{\psi }_L^\alpha \,\,\,\bar{\psi }_R^\alpha \right)
 \\ \label{whatischi}
 \chi^a_\alpha =
 \begin{pmatrix}
   \chi ^a_{L\,\alpha }   \\
   \chi ^a_{R\,\alpha }  \\
 \end{pmatrix},&\qquad &
 \bar{\chi }^\alpha _a=-i\varepsilon ^{\alpha \beta }\varepsilon _{ab}
 \left(-\chi ^b_{R\,\beta }\,\,\,\chi^b_{L\,\beta } \right)
 =-i\varepsilon ^{\alpha \beta }\varepsilon _{ab}\chi
 ^{b\,T}_\beta \gamma ^1.
\end{eqnarray}
The Dirac matrices $\gamma ^\mu $ are chosen to be
\begin{equation}\label{}
 \gamma ^\mu =(\sigma ^1,i\sigma ^2).
\end{equation}
The symbol $\Pi_\mu ^{\nu \lambda }$ is defined as
\begin{equation}\label{}
 \Pi_\mu ^{\nu \lambda }=\delta _\mu ^0h^{\nu \lambda
 }+\frac{1}{\sqrt{-h}}\,\delta ^1_\mu \epsilon ^{\nu \lambda }.
\end{equation}

Fixing the light-cone gauge at the quadratic level in fluctuations
amounts to treating $t$ and $\varphi $ as classical background
fields: $t=\tau =\varphi $ and replacing $h^{\mu \nu }$ by the flat
Minkowski metric $\eta_{\mu \nu }$. Then $\partial _\mu t=\delta
^0_\mu =\partial _\mu \varphi $, $\Pi^{\nu 0}_\mu=\delta ^\nu _\mu
$, and (\ref{L2}) becomes
\begin{eqnarray}\label{L2gf}
 \mathcal{L}^{(2)}_{\rm g.f.}&=&
 \partial _\mu \bar{X}_a\partial ^\mu X^a-\frac{1}{4}\,\bar{X}_aX^a
 +\frac{1}{2}\left(\partial _\mu Z\right)^2-\frac{1}{2}\,Z^2
 +\frac{1}{2}\left(\partial _\mu Y_i\right)^2-\frac{1}{2}\,Y^2_i
 \nonumber \\ &&
 +i\bar{\psi }^\alpha\gamma ^\mu\partial _\mu \psi _\alpha
 -\frac{1}{2}\bar{\psi }^\alpha\psi _\alpha
 +\frac{i}{2}\,\bar{\chi }^\alpha _a\gamma ^\mu \partial
 _\mu\chi ^a_\alpha-\frac{1}{2}\,\bar{\chi }^\alpha _a \chi
 ^a_\alpha.
\end{eqnarray}
This Lagrangian describes two complex scalars and two Dirac fermions
of mass $1/2$, and four real scalars and four Majorana fermions of
mass $1$ (the second equation in (\ref{whatischi}) can be regarded
as a Majorana condition). The Lagrangian is equivalent to the one
derived in \cite{Arutyunov:2008if}, although we have used a
different coset parameterization. The light modes $(X^a,\psi _\alpha
)$, $(\bar{X}_a,\bar{\psi }^\alpha )$ correspond to particles and
anti-particles of the worldsheet scattering theory and transform in
the $(2|2)_{\pm 1}$ representations of the unbroken symmetry group
$PSU(2|2)\times U(1)$. The heavy modes $(Z,Y_i,\chi ^a_\alpha )$ are
neutral under $U(1)$ and transform in the $(\mathbf{1},\mathbf{1})$,
$(\mathbf{1},\mathbf{3})$ and $(\mathbf{2},\mathbf{2})$
representations of $SU(2)\times SU(2)$.

Higher orders of the expansion in $\mathbb{X}$ describe interactions
of the BMN modes, which will lead to non-trivial scattering and to
quantum corrections. Our goal will be to compute the imaginary part
of the polarization operator for the $Z$ field at one loop and the
$2\rightarrow 2$ scattering amplitudes for $X^a$ and $\bar{X}_a$ at
the tree level. For that we need to expand the sigma-model
Lagrangian to the quartic order in fluctuations. The resulting
general expressions are quite complicated, but for this particular
computation only a few terms are necessary. These terms in fact are
quite simple. At the cubic level, the interaction terms that involve
$Z$ and two light fields are generally of the form $Z\bar{\psi }\psi
$ and $Z\bar{X}X$. In the parameterization (\ref{cosetX}) the
$Z\bar{\psi }\psi $ vertex is absent, and we are left with the
unique $Z\bar{X}X$ term:
\begin{equation}\label{L3}
 \mathcal{L}^{(3)}=i\,Z\bar{X}_a\dder{\mu} X^a\partial
 ^\mu \varphi +\ldots .
\end{equation}
Again, the gauge fixing amounts to replacing $\partial _\mu \varphi
$ by $\delta _\mu ^0$ and $h^{\mu \nu }$ by the Minkowski metric
(see appendix~\ref{lcg} for a more rigorous argument):
\begin{equation}\label{L3gf}
 \mathcal{L}^{(3)}_{\rm g.f.}=iZ\bar{X}_a\dder{0}X^a+\ldots .
\end{equation}

At the quartic level the only terms we need are the $X$, $\bar{X}$
self-interactions:
\begin{eqnarray}\label{L4}
 \mathcal{L}^{(4)}&=&\frac{1}{24}\left(\bar{X}_aX^a\right)^2\left(\partial _\mu \varphi
 \right)^2+\frac{1}{6}\left(\partial _\mu \bar{X}_aX^a\right)^2+
 \frac{1}{6}\left(\bar{X}_a\partial _\mu X^a\right)^2
 -\frac{1}{6}\,\bar{X}_a\partial _\mu X^a\partial ^\mu \bar{X}_bX^b
 \nonumber \\ &&
 -\frac{1}{6}\,\partial _\mu \bar{X}_a\partial ^\mu X^a\bar{X}_bX^b+\ldots.
\end{eqnarray}
Applying the gauge-fixing procedure outlined in  appendix~\ref{lcg}
to the Lagrangian (\ref{L0}), (\ref{L2}), (\ref{L3}), (\ref{L4}) we
get
\begin{eqnarray}\label{L4gf}
 \mathcal{L}^{(4)}_{\rm g.f.}&=&
 \frac{6a-5}{96}\left(\bar{X}_aX^a\right)^2+\frac{1}{6}\left(\partial _\mu \bar{X}_aX^a\right)^2+
 \frac{1}{6}\left(\bar{X}_a\partial _\mu X^a\right)^2-\frac{1}{6}\,\bar{X}_a\partial _\mu X^a\partial ^\mu \bar{X}_bX^b
 \nonumber \\ &&
 -\frac{1}{6}\,\partial _\mu \bar{X}_a\partial ^\mu X^a\bar{X}_bX^b
 -\frac{1-2a}{2}\left(\partial _\mu \bar{X}_a\partial ^\mu
 X^a\right)^2
 \nonumber \\ &&
 +\frac{1-2a}{4}\left(\partial _\mu \bar{X}_a\partial _\nu X^a
 +\partial _\nu \bar{X}_a\partial _\mu X^a\right)^2
 +\frac{1}{4}\,\bar{X}_aX^a\left(\partial _0\bar{X}_b\partial _0X^b
 +\partial _1\bar{X}_b\partial _1X^b\right)+\ldots\!,
\end{eqnarray}
where $a$ is the gauge-fixing parameter, $0\leq a\leq 1/2$, which
interpolates between the temporal gauge ($a=0$) and the pure
light-cone gauge ($a=1/2$) \cite{Arutyunov:2006gs}. The gauge-fixed
string action is
\begin{equation}\label{lcsa}
 S_{\rm l.c.}=\frac{\sqrt{2\lambda}}{2}\int_{}^{}d^2x\,
 \left(\mathcal{L}^{(2)}_{\rm g.f.}+\mathcal{L}^{(3)}_{\rm g.f.}+\mathcal{L}^{(4)}_{\rm g.f.}
 +\ldots \right),
\end{equation}
where the fields are periodic in $x^1$ with the period
$\mathcal{J}_+=2[(1-a)J+aE]/\sqrt{2\lambda }$ (appendix~\ref{lcg}).
We will ignore this periodicity by sending $\mathcal{J}_+$ to
infinity, and will treat (\ref{lcsa}) as a massive two-dimensional
field theory on  a plane with the coupling constant
$\sqrt{2\lambda}/2$.

Our gauge-fixed Lagrangian is different from those in
\cite{Astolfi:2008ji} and \cite{Sundin:2008vt}, where different sets
of coordinates on $CP^3$ were used. The Lagrangians should be
related by field redefinitions. In the next two sections we will use
the light-cone string Lagrangian to compute the one-loop
polarization operator for $Z$, and the tree level scattering matrix
for $X$, $\bar{X}$.

\section{Quantum corrections for the heavy mode}

\begin{figure}[t]
\centerline{\includegraphics[width=8cm]{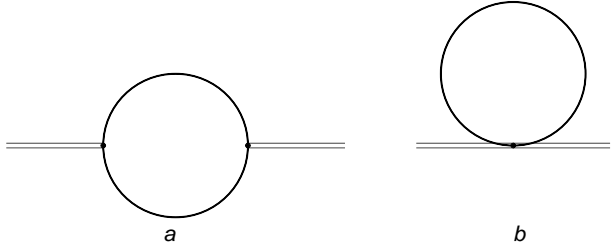}}
\caption{\label{selfe}\small The one-loop self-energy graphs.}
\end{figure}

There are two types of diagrams that contribute to the self-energy
of $Z$ at one loop (fig.~\ref{selfe}). In principle all cubic and
quartic vertices that involve $Z$ are necessary to compute the
one-loop polarization operator, but only the heavy-light-light
vertex contributes to the threshold singularity at $\omega =E(p)$ we
are interested in. Indeed, the bubble graph, fig.~\ref{selfe}b, does
not have an imaginary part, and the two-particle cut of the
diagram~\ref{selfe}a with a heavy mode in the loop starts at $\omega
=2E(p/2)$ rather than at $\omega =E(p)=2\varepsilon (p/2)$.

Since $Z$ couples to the charge density of $X$, the polarization
operator $\Pi(\omega ,p)$  can be expressed through the $00$
component of the correlator of currents:
\begin{equation}\label{}
 \Pi=\frac{2}{\sqrt{2\lambda }}\,\Pi_{00},
\end{equation}
where
\begin{equation}\label{}
 \Pi_{\mu \nu }=i\int_{}^{}d^2x\,\,{\rm e}\,^{-i\vec{p}\vec{x}}\left\langle
 J_\mu (x)J_\nu (0)\right\rangle, \qquad
 J_\mu =i\bar{X}_a\dder{\mu }X^a.
\end{equation}
The standard Feynman parameter representation for the loop integral
gives
\begin{equation}\label{}
 \Pi_{\mu \nu }=\frac{1}{8\pi }\left\{\frac{2}{\varepsilon }-\gamma
 +\ln(4\pi )
 +\int_{0}^{1}dx\,
 \left[
 2\eta_{\mu \nu }\ln\left(\frac{1}{4}-x(1-x)\vec{p}^{\,2}\right)+
 \frac{(1-2x)^2p_\mu p_\nu }{\frac{1}{4}-x(1-x)\vec{p}^{\,2}}\right]
 \right\},
\end{equation}
where we have used dimensional regularization to cut off the
divergence. The divergent term does not contribute to
$\mathop{\mathrm{Im}}\Pi$ and should cancel against the bubble
graph~\ref{selfe}b plus the heavy mode loop in~\ref{selfe}a.

Near the threshold ($\vec{p}^{\,2}=1$)\footnote{Here
$\mathop{\mathrm{Im}}\Pi$ denotes the non-analytic part of the
polarization operator: $\Pi=\mathop{\mathrm{Im}}\Pi+{\rm analytic}$.
$\mathop{\mathrm{Im}}\Pi$ is real below the threshold and becomes
pure imaginary above the threshold. In the equations below we assume
that $\vec{p}^2<1$ and $\omega <E(p)$. The sign of the polarization
operator in this region determines the fate of the single-particle
pole, as discussed in the introduction.}:
\begin{equation}\label{}
 \mathop{\mathrm{Im}}\Pi_{\mu \nu }\simeq-\frac{1}{4}\,\left(p_\mu p_\nu
 -\eta_{\mu \nu }\right)\sqrt{1-\vec{p}^{\,2}}\,,
\end{equation}
and
\begin{equation}\label{}
 \mathop{\mathrm{Im}}
 \Pi(\omega ,p)\simeq-\frac{\sqrt{2E(p)}}{2\sqrt{2\lambda }}\,p^2\sqrt{E(p)-\omega
 }\,.
\end{equation}
The function $C$ in (\ref{threshprop}) thus is given by
\begin{equation}\label{}
 C(p)=-\frac{p^2}{2\sqrt{2E(p)}}\,,
\end{equation}
and is strictly negative. From the discussion following
eq.~(\ref{threshprop}) we conclude that the one-particle pole in the
$Z$ propagator disappears once quantum corrections are taken into
account. There is no world-sheet particle associated with $Z$, it
dissolves in the $X$--$\bar{X}$ continuum. We expect the same to
happen with the other heavy modes, by supersymmetry. Indeed, there
are terms of the form $Y_i\bar{\psi }^\alpha \sigma _{i\alpha}
^{\hphantom{i\alpha }\beta }\psi _\beta $ and $\bar{\chi }^\alpha
_a\psi _\alpha \bar{X}_b\varepsilon ^{ab}$ in the cubic Lagrangian
that can mediate the mixing of $Y_i$ and $\chi _\alpha ^a$ with the
light modes. Strictly speaking the conclusion that the heavy modes
dissolve in the continuum also requires that the real part of the
polarization operator vanishes at the threshold. This we have not
checked, but it is plausible that supersymmetry cancelations
guarantee the absence of mass renormalization, because the heavy
modes lie in the semi-short representation of $PSU(2|2)$ and should
be BPS protected.

\section{Worldsheet scattering}

\begin{figure}[t]
\centerline{\includegraphics[width=14cm]{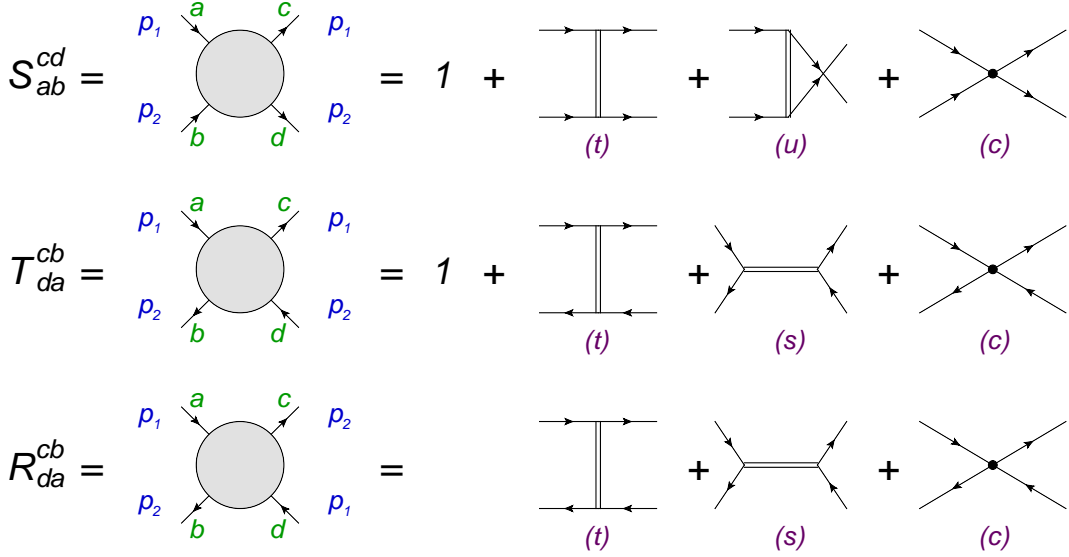}}
\caption{\label{scatmat}\small The scattering ($XX\rightarrow XX$),
transmission ($X\bar{X}\rightarrow X\bar{X}$) and reflection
($X\bar{X}\rightarrow \bar{X}X$) amplitudes. At tree level, they are
given by the sum of the contact interaction diagram (c) and the
exchange graphs with the heavy mode $Z$ in the (s), (t), and (u)
channels.}
\end{figure}

The only physical excitations on the string worldsheet are the
$(2|2)+(2|2)$ light modes $X^a$ and $\psi _\alpha $. Their
scattering matrix, whose form was conjectured in \cite{Ahn:2008aa},
underlies the exact solution of the model. Here we will compute the
worldsheet scattering amplitudes to the first order in
$2/\sqrt{2\lambda }$ and compare them with the strong-coupling
expansion of the conjectured exact S-matrix. For simplicity we
consider only the $XX\rightarrow XX$ and $X\bar{X}\rightarrow
X\bar{X}$ amplitudes. In this case there are three channels, defined
in fig.~\ref{scatmat}: the scattering $\mathbbm{S}_{ab}^{cd}$, the
transmission $\mathbbm{T}_{da}^{cb}$ and the reflection
$\mathbbm{R}_{da}^{cb}$.

\subsection{Worldsheet calculation}

At tree level,
\begin{eqnarray}\label{exapnd}
 \mathbbm{S}_{ab}^{cd}&=&\delta _a^c\delta _b^d+\frac{2i}{\sqrt{2\lambda
 }}\,S_{ab}^{cd}+\ldots
 \nonumber \\
\mathbbm{T}_{da}^{cb}&=&\delta _a^c\delta
_d^b+\frac{2i}{\sqrt{2\lambda
 }}\,T_{da}^{cb}+\ldots
 \nonumber \\
\mathbbm{R}_{da}^{cb}&=&\frac{2i}{\sqrt{2\lambda
 }}\,R_{da}^{cb}+\ldots.
\end{eqnarray}
The diagrams that contribute to $S_{ab}^{cd}$, $T_{da}^{cb}$ and
$R_{da}^{cb}$ are shown in fig.~\ref{scatmat}. It is straightforward
to compute these diagrams from the vertices in the gauge-fixed
Lagrangian (\ref{L2gf}), (\ref{L3gf}), (\ref{L4gf}). In order to get
the S-matrix elements one needs to take into account the Jacobian
from the momentum-conservation delta-function:
\begin{equation}\label{}
 \delta ^{(2)}\left(p_1^\mu +p_2^\mu -k_1^\mu -k_2^\mu \right)
 =\frac{\varepsilon _1\varepsilon _2}
 {\varepsilon _1p_2-\varepsilon _2p_1}\left[
 \delta (p_1-k_1)\delta (p_2-k_2)-\delta (p_1-k_2)\delta (p_2-k_1)
 \right],
\end{equation}
and the external leg factor $1/(4\varepsilon_1 \varepsilon_2 )$.
Collecting various pieces together, we get:
\begin{eqnarray}\label{Stree}
 S_{ab}^{cd}&=&\frac{1}{4}\,\,\frac{1}{\varepsilon _1p_2-\varepsilon _2p_1}
 \left\{
 \left[
 \frac{1}{2}\left((p_1+p_2)^2-2\varepsilon _1\varepsilon _2
 +4(1-2a)(\varepsilon _1p_2-\varepsilon _2p_1)^2\right)_{(c)}
 +\varepsilon _1\varepsilon _2\,{}_{(t)}\right]
 \right. \nonumber \\ &&\times\left.
 \delta _a^c\delta _b^d
 +\left[
 \frac{1}{8}\left(12p_1p_2-4\varepsilon _1\varepsilon
_2-1\right)_{(c)}
 +\frac{1}{8}\left(4p_1p_2+4\varepsilon _1\varepsilon _2+1\right)_{(u)}
 \right]\delta _a^d\delta _b^c
 \right\}
 \\
  T_{da}^{cb}&=&\frac{1}{4}\,\,\frac{1}{\varepsilon _1p_2-\varepsilon _2p_1}
 \left\{
 \left[
 \frac{1}{2}\left((p_1-p_2)^2+2\varepsilon _1\varepsilon _2
 +4(1-2a)(\varepsilon _1p_2-\varepsilon _2p_1)\right)_{(c)}
 -\varepsilon_1\varepsilon _2\,{} _{(t)}
 \right]
 \right. \nonumber \\ &&\times \left.
 \delta _a^c\delta _d^b
 +\left[\frac{1}{8}\left(4\varepsilon _1\varepsilon
 _2-12p_1p_2-1\right)_{(c)}
 -\frac{1}{8}\left(4\varepsilon _1\varepsilon _2+4p_1p_2-1\right)_{(s)}
 \right]
 \delta _a^b\delta _d^c
 \right\}
 \\
 R_{da}^{cb}&=&\frac{1}{4}\,\,\frac{1}{\varepsilon _1p_2-\varepsilon_2p_1}
 \left\{
 \left[\frac{1}{8}
 \left(4\varepsilon _1\varepsilon _2+p_1p_2+1\right)_{(c)}
 -\frac{1}{8}
 \left(4\varepsilon _1\varepsilon _2+p_1p_2+1\right)_{(t)}
 \right]\delta _a^c\delta _d^b
 \right. \nonumber \\ &&\left.
 +\left[\frac{1}{8}
 \left(-4\varepsilon _1\varepsilon _2-p_1p_2+1\right)_{(c)}
 +\frac{1}{8}
 \left(4\varepsilon _1\varepsilon _2+p_1p_2-1\right)_{(s)}
  \right]\delta _a^b\delta _d^c
 \right\},
\end{eqnarray}
The contribution from the exchange diagrams is structurally similar
to that of the contact interaction, and in fact the two
contributions tend to cancel. In the case of the reflection
amplitude the cancelation is complete:
\begin{eqnarray}\label{worldsheetSa}
 S_{ab}^{cd}&=&\left[
 \frac{1}{8}\,\,\frac{(p_1+p_2)^2}{\varepsilon _1p_2-\varepsilon _2p_1}
 +\frac{1}{2}\left(1-2a\right)\left(\varepsilon _1p_2-\varepsilon _2p_1\right)
 \right]\delta _a^c\delta _b^d
 +\frac{1}{2}\,\,\frac{p_1p_2}{\varepsilon _1p_2-\varepsilon
 _2p_1}\,\delta _a^d\delta _b^c
 \\
 T_{da}^{cb}&=&\left[
 \frac{1}{8}\,\,\frac{(p_1-p_2)^2}{\varepsilon _1p_2-\varepsilon _2p_1}
 +\frac{1}{2}\left(1-2a\right)\left(\varepsilon _1p_2-\varepsilon _2p_1\right)
 \right]\delta _a^c\delta _d^b
 -\frac{1}{2}\,\,\frac{p_1p_2}{\varepsilon _1p_2-\varepsilon
 _2p_1}\,\delta _a^b\delta _d^c
 \\
 R_{da}^{cb}&=&0.
\end{eqnarray}
The tree-level worldsheet S-matrix is indeed reflectionless, in
accord with the conjectured absence of reflection in the exact
soliton-antisoliton scattering \cite{Ahn:2008aa}. Let us check that
the transmission and scattering amplitudes also agree with the
S-matrix proposed in \cite{Ahn:2008aa}.

\subsection{Comparison to the exact result}

The exact S-matrix is expressed in terms of the kinematic variables
\begin{equation}\label{}
 x^\pm=\frac{1+\sqrt{1+16h^2(\lambda )\sin^2\frac{p}{\sqrt{2\lambda
 }}}}
 {4h(\lambda)\sin\frac{p}{\sqrt{2\lambda }}}\,
 \,{\rm e}\,^{\pm\frac{ip}{\sqrt{2\lambda }}},
\end{equation}
which at strong coupling become
\begin{equation}\label{}
 x^\pm\simeq
 \frac{1+2\varepsilon }{2p}\left(1\pm\frac{ip}{\sqrt{2\lambda
 }}\right).
\end{equation}

The exact S-matrix of $AdS_4/CFT_3$ is the Beisert's $PSU(2|2)$
S-matrix  \cite{Beisert:2005tm,Beisert:2006qh} with an appropriate
dressing phase. The $X^aX^b\rightarrow X^cX^d$ amplitude coincides
with the S-matrix element for the scalar states of the $(2|2)$
multiplet:
\begin{equation}\label{exactSm}
 \mathbb{S}_{ab}^{cd}=\,{\rm e}\,^{i\theta }\,
 \frac{1-\frac{1}{x_1^+x_2^-}}{1-\frac{1}{x_1^+x_2^+}}
 \,\,\frac{x_1^--x_2^-}{x_1^--x_2^+}\,\delta _a^c\delta _b^d
 +\,{\rm e}\,^{i\theta }\left(
 \frac{1-\frac{1}{x_1^+x_2^-}}{1-\frac{1}{x_1^-x_2^+}}
 \,\,\frac{x_1^+-x_2^-}{x_1^--x_2^+}
 -
 \frac{1-\frac{1}{x_1^+x_2^-}}{1-\frac{1}{x_1^+x_2^+}}
 \,\,\frac{x_1^--x_2^-}{x_1^--x_2^+}
 \right)\delta _a^d\delta _b^c,
\end{equation}
The dressing phase $\theta $ is mildly gauge-dependent
\cite{Arutyunov:2009ga}, and has the following general form:
\begin{equation}\label{}
 \theta =\frac{1}{\sqrt{2\lambda }}\left[
 p_2-p_1-2a(\varepsilon _1p_2-\varepsilon _2p_1)\right]
 -\frac{\sqrt{2\lambda }}{2}\sum_{r,s=\pm}^{}rs\chi (x_1^r,x_2^s).
\end{equation}
where \cite{Arutyunov:2004vx}
\begin{equation}\label{AFSphase}
 \chi (x,y)=(x-y)\left[\frac{1}{xy}+\left(1-\frac{1}{xy}\right)
 \ln\left(1-\frac{1}{xy}\right)\right]+O\left(\frac{1}{\sqrt{2\lambda
 }}\right).
\end{equation}
Higher orders in the dressing phase are known exactly
\cite{Beisert:2006ib}, and in fact non-perturbatively in $\lambda $
\cite{Beisert:2006ez,Dorey:2007xn}, but we will only need the
leading term explicitly shown in (\ref{AFSphase}).

Expanding (\ref{exactSm}) in $1/\sqrt{2\lambda }$, we get:
\begin{equation}\label{Sex}
 \mathbb{S}_{ab}^{cd}\simeq
 \left\{1+\frac{i}{\sqrt{2\lambda }}
 \left[
 \frac{1}{4}\,\,\frac{(p_1+p_2)^2}{\varepsilon _1p_2-\varepsilon _2p_1}
 +\left(1-2a\right)\left(\varepsilon _1p_2-\varepsilon _2p_1\right)
 \right]
 \right\}\delta _a^c\delta _b^d
 +\frac{i}{\sqrt{2\lambda }}\,\,\frac{p_1p_2}{\varepsilon _1p_2-\varepsilon
 _2p_1}\,\delta _a^d\delta _b^c,
\end{equation}
which agrees precisely with the explicit worldsheet calculation
(\ref{exapnd}), (\ref{worldsheetSa}).

To compare the $X^a\bar{X}_b\rightarrow X^c\bar{X}_d$ transmission
amplitude with the exact S-matrix of Ahn and Nepomechie
\cite{Ahn:2008aa}, we should take into account that in
\cite{Ahn:2008aa} the S-matrix is expressed in terms of the fields
with the upper $su(2)$ indices: $B^a=\varepsilon ^{ab}\bar{X}_b$.
The transmission amplitude is again given by the Beisert's
$PSU(2|2)$ S-matrix, with an addition dressing factor. The
transmission matrix $\mathbb{T}_{da}^{cb}$ defined in
fig.~\ref{scatmat} is then proportional to the $XX\rightarrow XX$
scattering matrix:
\begin{equation}\label{}
 \mathbb{T}_{da}^{cb}=S_0\varepsilon ^{bf}\varepsilon
 _{de}\mathbb{S}_{af}^{ce},
\end{equation}
where \cite{Ahn:2008aa}
\begin{equation}\label{}
 S_0=\frac{1-\frac{1}{x_1^-x_2^+}}{1-\frac{1}{x_1^+x_2^-}}\,\,
 \frac{x_1^--x_2^+}{x_1^+-x_2^-}\,.
\end{equation}
If we write
\begin{equation}\label{}
 \mathbb{S}_{ab}^{cd}=A\delta _a^c\delta _b^d+B\delta _a^d\delta
 _b^c,
\end{equation}
then
\begin{equation}\label{TthruS}
 \mathbb{T}_{da}^{cb}=S_0\left(A+B\right)\delta _a^c\delta _d^b-S_0B\delta
 _a^b\delta _d^c.
\end{equation}
The additional factor $S_0$ expands at strong coupling as
\begin{equation}\label{}
 S_0\simeq 1-\frac{i}{\sqrt{2\lambda }}\,\,\frac{2p_1p_2}{\varepsilon _1p_2-\varepsilon
 _2p_1}\,,
\end{equation}
and from (\ref{TthruS}), (\ref{Sex}) we find:
\begin{equation}\label{Tex}
 \mathbb{T}_{da}^{cb}\simeq
 \left\{1+\frac{i}{\sqrt{2\lambda }}
 \left[
 \frac{1}{4}\,\frac{(p_1-p_2)^2}{\varepsilon _1p_2-\varepsilon _2p_1}
 +\left(1-2a\right)\left(\varepsilon _1p_2-\varepsilon _2p_1\right)
 \right]
 \right\}\delta _a^c\delta _d^b
 -\frac{i}{\sqrt{2\lambda }}\,\frac{p_1p_2}{\varepsilon _1p_2-\varepsilon
 _2p_1}\,\delta _a^b\delta _d^c,
\end{equation}
again in the precise agreement with the direct worldsheet
calculation.

\section{Conclusions}

In this paper we studied the fate of the heavy BMN modes in type IIA
string theory on $AdS_4\times CP^3$. We argued that once the quantum
corrections are taken into account the heavy modes mix with the
two-particle continuum of the light modes and disappear from the
spectrum. This picture is confirmed by an explicit one-loop
calculation in the worldsheet sigma-model and is consistent with the
Bethe ansatz solution of the $AdS_4/CFT_3$ system, where heavy modes
were interpreted as stacks of Bethe roots. The true physical
excitations on the string worldsheet are the $4+4$ light modes. It
is interesting that all the bosonic light modes are fluctuations in
the $CP^3$ directions. The transverse modes in $AdS_4$ disappear by
mixing with the two-fermion states. This is in a qualitative
agreement with the spin-chain picture of the spectrum in the dual
$\mathcal{N}=6$ Chern-Simons theory. The AdS directions are dual to
the covariant derivatives in the field-theory operators. In the
Chern-Simons spin chain, the derivative operators mix with fermions:
$\Phi D_\mu\Phi\leftrightarrow \bar{\Psi }_\alpha \sigma _\mu
^{\hphantom{\mu }\alpha \beta }\Psi _\beta  $, and do not form a
closed sector \cite{Minahan:2008hf} \footnote{It is still possible
to define a closed $sl(2)$ sector \cite{Zwiebel:2009vb}, but this
sector has a half-fermionic ground state and is very far from the
dual of the BMN ground state in the light-cone sigma-model.}.

Despite that the heavy modes disappear from the spectrum, the
corresponding fields appear as intermediate states in the
perturbative worldsheet S-matrix\footnote{The S-matrix does not have
poles that correspond to the heavy modes (I would like to thank
T.~Harmark and M.~Orselli for the discussion of this point). They
just appear as internal lines in the Feynman diagrams for the
worldsheet scattering amplitudes.}. In fact, the exchange of the
heavy modes is crucially important for cancelation of the reflection
amplitude in the particle-antiparticle scattering.  We also checked
that the scattering and transmission amplitudes computed in the
light-cone sigma-model agree with the large-$\lambda $ expansion of
the exact S-matrix proposed in \cite{Ahn:2008aa}.

\subsection*{Acknowledgments}
I would like to thank C.~Ahn, G.~Arutyunov, S.~Frolov, S.~Hirano,
D.~Hofman, J.~Maldacena, J.~Minahan, S.-J.~Rey, P.~Vieira and
especially T.~Harmark and M.~Orselli for interesting discussions. I
also thank the organizers of the the 29th winter school "Geometry
and Physics" in Srni and of the workshop "Gauge Fields, Cosmology
and Mathematical String Theory" in Banff for hospitality during the
course of this work. This work was supported in part by the Swedish
Research Council under the contract 621-2007-4177, in part by the
RFFI grant 09-02-00253, and in part by the grant for support of
scientific schools NSH-3036.2008.2.

\appendix

\section{The $osp(6|4)$ superalgebra}\label{osp64}{\allowdisplaybreaks

The (anti)commutation relations of $osp(6|4)$ in the basis
(\ref{basis}) are
\begin{eqnarray}\label{}
 &&
 [R^a_b,R^c_d]=\delta ^a_dR^c_b-\delta ^c_bR^a_d,\qquad
 [R^a_b,R^c]=\delta ^a_bR^c-\delta ^c_bR^a,\qquad
 [R^a_b,R_c]=-\delta ^a_bR_c+\delta ^a_cR_b,
 \nonumber \\ &&
 [R^a,R_b]=\delta ^a_b\left(R^c_c-R\right)-R^a_b,\qquad
 [R,R^a]=-R^a,\qquad
 [R,R_a]=R_a,
 \nonumber \\ &&
 [T_i,T_j]=\varepsilon _{ijk}T_k,\qquad
 [T_i,K_j]=\varepsilon _{ijk}K_k,\qquad
 [K_i,K_j]=-\varepsilon _{ijk}T_k,
 \nonumber \\ &&
 [R^a_b,B^c]=-\delta ^c_bB^a,\qquad
 [R^a_b,B_c]=\delta ^a_cB_b,\qquad
 [R^a,B_b]=\frac{1}{2}\,\delta ^a_b\left(J+M\right),
 \nonumber \\ &&
 [R_a,B^b]=\frac{1}{2}\,\delta ^b_a\left(J-M\right),\qquad
 [R,B^a]=-B^a,\qquad
 [R,B_a]=B_a,
 \nonumber \\ &&
 [J,R^a_b]=-\delta ^a_bM,\qquad
 [J,R^a]=-B^a,\qquad
 [J,R_a]=-B_a,
 \nonumber \\ &&
 [M,R^a_b]=-\delta ^a_bJ,\qquad
 [M,R^a]=B^a,\qquad
 [M,R_a]=-B_a,
 \nonumber \\ &&
 [T_i,L_j]=\varepsilon _{ijk}L_k,\qquad
 [K_i,L_j]=-\delta _{ij}D,\qquad
 [D,K_i]=L_i,
 \nonumber \\ &&
 [B^a,B_b]=\delta ^a_bR+R^a_b,\qquad
 [J,B^a]=-R^a,\qquad
 [J,B_a]=-R_a,\qquad
 [J,M]=-2R^a_a,
 \nonumber \\ &&
 [M,B^a]=-R^a,\qquad
 [M,B_a]=R_a,\qquad
 [D,L_i]=-K_i,\qquad
 [L_i,L_j]=-\varepsilon _{ijk}T_k,
 \nonumber \\ &&
 [R^a_b,Q^c_\alpha ]=-\delta ^c_bQ^a_\alpha ,\qquad
 [R^a_b,\bar{Q}_{c\alpha }]=\delta ^a_c\bar{Q}_{b\alpha },\qquad
 [R^a,Q^b_\alpha ]=-\varepsilon ^{ab}Q_\alpha ,
 \nonumber \\ &&
 [R^a,\bar{Q}_\alpha ]=\varepsilon ^{ab}\bar{Q}_{b\alpha },\qquad
 [R_a,\bar{Q}_{b\alpha }]=\varepsilon _{ab}\bar{Q}_\alpha ,\qquad
 [R_a,Q_\alpha ]=-\varepsilon _{ab}Q^b_\alpha ,
 \nonumber \\ &&
 [R,Q_\alpha ]=-Q_\alpha ,\qquad
 [R,\bar{Q}_\alpha ]=\bar{Q}_\alpha ,\qquad
 [K_i,Q^a_\alpha ]=\frac{1}{2}\,\sigma _{i\alpha }^{\hphantom{i\alpha }\beta
 }Q^a_\beta ,
 \nonumber \\ &&
 [K_i,Q_\alpha ]=\frac{1}{2}\,\sigma _{i\alpha }^{\hphantom{i\alpha }\beta
 }Q_\beta ,\qquad
 [K_i,\bar{Q}_{a\alpha }]=-\frac{1}{2}\,\sigma _{i\alpha }^{\hphantom{i\alpha }\beta
 }\bar{Q}_{a\beta },\qquad
 [K_i,\bar{Q}_{\alpha }]=-\frac{1}{2}\,\sigma _{i\alpha }^{\hphantom{i\alpha }\beta
 }\bar{Q}_{\beta },
 \nonumber \\ &&
 [T_i,Q^a_\alpha ]=\frac{i}{2}\,\sigma _{i\alpha }^{\hphantom{i\alpha }\beta
 }Q^a_\beta ,\qquad
 [T_i,Q_\alpha ]=\frac{i}{2}\,\sigma _{i\alpha }^{\hphantom{i\alpha }\beta
 }Q_\beta ,
 \nonumber \\ &&
 [T_i,\bar{Q}_{a\alpha }]=\frac{i}{2}\,\sigma _{i\alpha }^{\hphantom{i\alpha }\beta
 }\bar{Q}_{a\beta} ,\qquad
 [T_i,\bar{Q}_\alpha ]=\frac{i}{2}\,\sigma _{i\alpha }^{\hphantom{i\alpha }\beta
 }\bar{Q}_{\beta} ,
 \nonumber \\ &&
 [R^a_b,S_{c\alpha }]=\delta ^a_cS_{b\alpha },\qquad
 [R^a_b,\bar{S}^c_\alpha ]=-\delta ^c_b\bar{S}^a_\alpha ,\qquad
 [R^a,\bar{S}^b_\alpha ]=-\varepsilon ^{ab}\bar{S}_\alpha ,
 \nonumber \\ &&
 [R^a,S_\alpha ]=\varepsilon ^{ab}S_{b\alpha },\qquad
 [R_a,S_{b\alpha }]=\varepsilon _{ab}S_\alpha ,\qquad
 [R_a,\bar{S}_\alpha ]=-\varepsilon _{ab}\bar{S}^b_\alpha ,
 \nonumber \\ &&
 [R,S_\alpha ]=S_\alpha ,\qquad
 [R,\bar{S}_\alpha ]=-\bar{S}_\alpha ,\qquad
 [K_i,S_{a\alpha }]=\frac{1}{2}\,\sigma _{i\alpha }^{\hphantom{i\alpha }\beta
 }S_{a\beta },
  \nonumber \\ &&
 [K_i,S_\alpha ]=\frac{1}{2}\,\sigma _{i\alpha }^{\hphantom{i\alpha }\beta
 }S_{\beta },\qquad
 [K_i,\bar{S}^a_\alpha ]=-\frac{1}{2}\,\sigma _{i\alpha }^{\hphantom{i\alpha }\beta
 }\bar{S}^a_{\beta },\qquad
 [K_i,\bar{S}_\alpha ]=-\frac{1}{2}\,\sigma _{i\alpha }^{\hphantom{i\alpha }\beta
 }\bar{S}_{\beta },
 \nonumber \\ &&
 [T_i,S_{a\alpha }]=\frac{i}{2}\,\sigma _{i\alpha }^{\hphantom{i\alpha }\beta
 }S_{a\beta },\qquad
 [T_i,S_\alpha ]=\frac{i}{2}\,\sigma _{i\alpha }^{\hphantom{i\alpha }\beta
 }S_{\beta },
 \nonumber \\ &&
 [T_i,\bar{S}^a_\alpha ]=\frac{i}{2}\,\sigma _{i\alpha }^{\hphantom{i\alpha }\beta
 }\bar{S}^a_{\beta },\qquad
 [T_i,\bar{S}_\alpha ]=\frac{i}{2}\,\sigma _{i\alpha }^{\hphantom{i\alpha }\beta
 }\bar{S}_{\beta },
 \nonumber \\ &&
 [B^a,\bar{Q}_{b\alpha }]=\delta ^a_b\bar{S}_\alpha ,\qquad
 [B^a,\bar{Q}_\alpha ]=-\bar{S}^a_\alpha ,\qquad
 [B_a,Q^b_\alpha ]=\delta ^b_aS_\alpha ,\qquad
 [B_a,Q_\alpha ]=-S_{a\alpha },
 \nonumber \\ &&
 [J,Q^a_\alpha ]=\varepsilon ^{ab}S_{b\alpha },\qquad
 [J,\bar{Q}_{a\alpha }]=-\varepsilon _{ab}\bar{S}^b_\alpha ,\qquad
 [M,Q^a_\alpha ]=\varepsilon ^{ab}S_{b\alpha },
 \nonumber \\ &&
 [M,\bar{Q}_{a\alpha }]=\varepsilon _{ab}\bar{S}^b_\alpha ,\qquad
 [L_i,Q^a_\alpha ]=\frac{1}{2}\,\sigma _{i\alpha }^{\hphantom{i\alpha }\beta
 }\bar{S}^a_\beta ,\qquad
 [L_i,Q_\alpha ]=\frac{1}{2}\,\sigma _{i\alpha }^{\hphantom{i\alpha }\beta
 }\bar{S}_\beta ,
 \nonumber \\ &&
 [L_i,\bar{Q}_{a\alpha} ]=\frac{1}{2}\,\sigma _{i\alpha }^{\hphantom{i\alpha }\beta
 }S_{a\beta },\qquad
 [L_i,\bar{Q}_\alpha ]=\frac{1}{2}\,\sigma _{i\alpha }^{\hphantom{i\alpha }\beta
 }S_\beta ,\qquad
 [D,Q^a_\alpha ]=\frac{1}{2}\,\bar{S}^a_\alpha ,
 \nonumber \\ &&
 [D,Q_\alpha ]=\frac{1}{2}\,\bar{S}_\alpha ,\qquad
 [D,\bar{Q}_{a\alpha }]=-\frac{1}{2}\,S_{a\alpha },\qquad
 [D,\bar{Q}_\alpha ]=-\frac{1}{2}\,S_\alpha ,
 \nonumber \\ &&
 [B^a,S_{b\alpha }]=\delta ^a_bQ_\alpha ,\qquad
 [B^a,S_\alpha ]=-Q^a_\alpha ,\qquad
 [B_a\bar{S}^b_\alpha ]=\delta ^b_a\bar{Q}_\alpha ,\qquad
 [B_a,\bar{S}_\alpha ]=-\bar{Q}_{a\alpha },
 \nonumber \\ &&
 [J,\bar{S}^a_\alpha ]=\varepsilon ^{ab}\bar{Q}_{b\alpha },\qquad
 [J,S_{a\alpha }]=-\varepsilon _{ab} Q^b_\alpha ,\qquad
 [M,\bar{S}^a_\alpha ]=\varepsilon ^{ab}\bar{Q}_{b\alpha },
 \nonumber \\ &&
 [M,S_{a\alpha }]=\varepsilon _{ab}Q^b_\alpha ,\qquad
 [L_i,S_{a\alpha }]=\frac{1}{2}\,\sigma _{i\alpha }^{\hphantom{i\alpha }\beta
 }\bar{Q}_{a\beta },\qquad
 [L_i,S_\alpha ]=\frac{1}{2}\,\sigma _{i\alpha }^{\hphantom{i\alpha }\beta
 }\bar{Q}_{\beta },
 \nonumber \\ &&
 [L_i,\bar{S}^a_\alpha ]=\frac{1}{2}\,\sigma _{i\alpha }^{\hphantom{i\alpha }\beta
 }Q^a_{\beta },\qquad
 [L_i,\bar{S}_\alpha ]=\frac{1}{2}\,\sigma _{i\alpha }^{\hphantom{i\alpha }\beta
 }Q_{\beta },\qquad
 [D,S_{a\alpha }]=\frac{1}{2}\,\bar{Q}_{a\alpha },
 \nonumber \\ &&
 [D,S_\alpha ]=\frac{1}{2}\,\bar{Q}_\alpha ,\qquad
 [D,\bar{S}^a_\alpha ]=-\frac{1}{2}\,Q^a_\alpha ,\qquad
 [D,\bar{S}_\alpha ]=-\frac{1}{2}\,Q_\alpha ,
 \nonumber \\ &&
 \{Q^a_\alpha ,Q^b_\beta \}=-\frac{i}{2}\,\varepsilon _{\alpha \beta }\varepsilon
 ^{ab}\left(J-M\right),\qquad
 \{Q^a_\alpha ,Q_\beta \}=i\varepsilon _{\alpha \beta }B^a,
 \nonumber \\ &&
 \{Q^a_\alpha ,\bar{Q}_{b\beta }\}=-i\delta ^a_b
 \left[\varepsilon _{\alpha \beta }D-\left(\sigma_i\varepsilon \right)_{\alpha \beta
 }L_i\right],\qquad
 \{Q_\alpha ,\bar{Q}_\beta \}=-i\left[\varepsilon _{\alpha \beta }D
 -\left(\sigma_i\varepsilon \right)_{\alpha \beta
 }L_i\right],
 \nonumber \\ &&
 \{\bar{Q}_{a\alpha },\bar{Q}_{b\beta }\}=\frac{i}{2}\,\varepsilon
 _{\alpha \beta }\varepsilon _{ab}\left(J+M\right),\qquad
 \{\bar{Q}_{a\alpha },\bar{Q}_\beta \}=i\varepsilon _{\alpha \beta
 }B_a,
 \nonumber \\ &&
 \{S_{a\alpha },S_{b\beta }\}=\frac{i}{2}\,\varepsilon _{\alpha \beta }\varepsilon
 _{ab}\left(J+M\right),\qquad
 \{S_{a\alpha },S_\beta \}=i\varepsilon _{\alpha \beta }B_a,
 \nonumber \\ &&
 \{S_{a\alpha },\bar{S}^b_\beta \}=-i\delta ^b_a\left[\varepsilon _{\alpha \beta }D
 -\left(\sigma_i\varepsilon \right)_{\alpha \beta
 }L_i\right],\qquad
 \{S_\alpha ,\bar{S}_\beta \}=-i\left[\varepsilon _{\alpha \beta }D
 -\left(\sigma_i\varepsilon \right)_{\alpha \beta
 }L_i\right],
 \nonumber \\ &&
 \{\bar{S}^a_\alpha ,\bar{S}^b_\beta \}=-\frac{i}{2}\,\varepsilon
 _{\alpha \beta }\varepsilon ^{ab}\left(J-M\right),\qquad
 \{\bar{S}^a_\alpha ,\bar{S}_\beta \}=i\varepsilon _{\alpha \beta
 }B^a,
 \nonumber \\ &&
 \{Q^a_\alpha ,S_{b\beta }\}=i\delta ^a_b\left(\sigma_i\varepsilon
 \right)_{\alpha \beta }\left(K_i-iT_i\right)+i\varepsilon _{\alpha \beta
 }R^a_b,\qquad
 \{Q^a_\alpha ,S_\beta \}=-i\varepsilon _{\alpha \beta }\varepsilon
 ^{ab}R_b,
 \nonumber \\ &&
 \{Q_\alpha ,S_{a\beta }\}=-i\varepsilon _{\alpha \beta }\varepsilon
 _{ab}R^b,\qquad
 \{Q_\alpha ,S_\beta \}=i\left(\sigma_i\varepsilon
 \right)_{\alpha \beta }\left(K_i-iT_i\right)+i\varepsilon _{\alpha \beta
 }R,
 \nonumber \\ &&
 \{\bar{Q}_{a\alpha },\bar{S}^b_\beta \}=-i\delta ^b_a\left(\sigma_i\varepsilon
 \right)_{\alpha \beta }\left(K_i+iT_i\right)-i\varepsilon _{\alpha \beta
 }R^b_a,\qquad
 \{\bar{Q}_{a\alpha} ,\bar{S}_\beta \}=i\varepsilon _{\alpha \beta }\varepsilon
 _{ab}R^b,
 \nonumber \\ &&
 \{\bar{Q}_\alpha ,\bar{S}^a_\beta \}=i\varepsilon _{\alpha \beta }\varepsilon
 ^{ab}R_b,\qquad
 \{\bar{Q}_\alpha ,\bar{S}_\beta \}=-i\left(\sigma_i\varepsilon
 \right)_{\alpha \beta }\left(K_i+iT_i\right)-i\varepsilon _{\alpha \beta
 }R,
\end{eqnarray}
where $\sigma _{i\alpha }^{\hphantom{i\alpha }\beta }$ are Pauli
matrices.

The non-zero components of the invariant metric are
\begin{eqnarray}\label{}
 &&
 \mathop{\mathrm{Str}}R^a_bR^c_d=-\delta ^a_d\delta ^c_b,\qquad
 \mathop{\mathrm{Str}}R^aR_b=-\delta ^a_b,\qquad
 \mathop{\mathrm{Str}}R^2=-1,\qquad
 \mathop{\mathrm{Str}} B^aB_b=\delta ^a_b
 \nonumber \\ &&
 \mathop{\mathrm{Str}} J^2=-2,\qquad
 \mathop{\mathrm{Str}} M^2=2,\qquad
 \mathop{\mathrm{Str}} D^2=-\frac{1}{2}\,,\qquad
 \mathop{\mathrm{Str}} L_iL_j=\frac{1}{2}\,\delta _{ij},
 \nonumber \\ &&
 \mathop{\mathrm{Str}} K_iK_j=\frac{1}{2}\,\delta _{ij},\qquad
 \mathop{\mathrm{Str}} T_iT_j=-\frac{1}{2}\,\delta _{ij},\qquad
 \mathop{\mathrm{Str}} Q^a_\alpha S_{b\beta }=i\varepsilon _{\alpha \beta }\delta
 ^a_b,
 \nonumber \\ &&
 \mathop{\mathrm{Str}} Q_\alpha S_\beta =i\varepsilon _{\alpha \beta
 },\qquad
 \mathop{\mathrm{Str}} \bar{Q}_{a\alpha }\bar{S}^b_\beta= i\varepsilon
 _{\alpha \beta }\delta ^b_a,\qquad
 \mathop{\mathrm{Str}} \bar{Q}_\alpha \bar{S}_\beta =i\varepsilon
 _{\alpha \beta }.
\end{eqnarray}

}
\section{Light-cone gauge}\label{lcg}

In this appendix we consider the light-cone gauge fixing for the
bosonic string in the background
\begin{equation}\label{metric}
 ds^2=-G_{tt}dt^2+G_{\varphi\varphi}d\varphi
 ^2+2G_{\varphi i}d\varphi dx^i+G_{ij}dx^idx^j,
\end{equation}
where $G_{tt}$, $G_{\varphi \varphi }$, $G_{\varphi i}$ and $G_{ij}$
are functions of $x^i$. This is a slight generalization over
\cite{Klose:2006zd,Arutyunov:2009ga} where the case of $G_{\varphi
i}=0$ has been considered.

The light-cone gauge fixing can be done in three steps
\cite{Kruczenski:2004cn}: one can T-dualize in the $\varphi $
direction, integrate out the worldsheet metric and then fix the
gauge in the resulting Nambu-Goto action. This yields a
non-polynomial Lagrangian, which can be further expanded in the
transverse fluctuations. T-duality is equivalent to gauging the
isometry
\begin{equation}\label{iso}
\varphi \rightarrow \varphi -(1-a)\xi,\qquad t\rightarrow t+a\xi
\end{equation}
of the metric (\ref{metric}), and then imposing the flatness
condition on the gauge connection. Here $0\leq a\leq 1/2$ is a gauge
parameter that interpolated between the temporal ($a=0$) and
light-cone ($a=1/2$) gauges \cite{Arutyunov:2006gs}. The worldsheet
Lagrangian becomes
\begin{eqnarray}\label{sdirect}
 \mathcal{L}&=&
 -\frac{G_{tt}}{2}\left(\partial _\mu t+aA_\mu \right)^2
 +\frac{G_{\varphi\varphi}}{2}\left[
 \partial _\mu \varphi -\left(1-a\right)A_\mu
 \right]^2
 \nonumber \\ &&
 +G_{\varphi i}\left[
 \partial _\mu \varphi -\left(1-a\right)A_\mu
 \right]\partial ^\mu X^i
 +\frac{G_{ij}}{2}\,\partial _\mu X^i\partial ^\mu X^j
 +\frac{1}{\sqrt{-h}}\,\tilde{\varphi } \varepsilon ^{\mu \nu }\partial
 _\mu A_\nu,
\end{eqnarray}
where $\tilde{\varphi }$ is a Lagrange multiplier that imposes the
flatness condition on $A_\mu $. The equations of motion for $A_\mu $
imply that
\begin{equation}\label{}
 \varepsilon ^{\mu \nu }\partial _{\nu }\tilde{\varphi }=\frac{J^\mu }{T }\,,
\end{equation}
where $J^\mu $ is the Neother current associated with the isometry
(\ref{iso}) and $T$ is the string tension ($T=\sqrt{2\lambda }/2$ in
$AdS_4\times CP^3$). Setting $\mu =0$ and integrating along the
fixed-time section of the string worldsheet, we find:
\begin{equation}\label{bc}
 \tilde{\varphi }(x^0,x^1+L)-\tilde{\varphi }(x^0,x^1)
 =\int_{-L/2}^{L/2}dx^1\,\partial _1\tilde{\varphi }
 =\mathcal{J}_+\equiv\frac{J_+}{T}\,,
\end{equation}
where $J_+$ is the corresponding Neother charge ($J_+=(1-a)J+aE$,
where $J$ is the angular momentum of the string and $E$ is its
energy), and $L$ is the internal length of the string that can be
chosen arbitrarily due to the reparameterization invariance.

The T-dual Lagrangian arises after integrating out $A_\mu $ in
(\ref{sdirect}):
\begin{eqnarray}\label{}
 \tilde{\mathcal{L}}&=&\frac{1}{2}\,\,
 \frac{1}{\left(1-a\right)^2G_{\varphi\varphi}-a^2G_{tt}}
 \left\{
 -G_{tt}G_{\varphi\varphi}
 \left(\partial _\mu X^+\right)^2+\left(\partial _\mu \tilde{\varphi }\right)^2
 -2aG_{tt}G_{\varphi i}\partial _\mu X^+\partial ^\mu X^i
 \vphantom{\frac{a}{1-a}}\right. \nonumber \\ &&\left.
 +\tilde{G}_{ij}\partial _\mu X^i\partial ^\mu X^j
 -\frac{1}{\sqrt{-h}}\,\varepsilon ^{\mu \nu }\partial _\mu \tilde{\varphi }
 \left[\frac{a}{1-a}G_{tt}\partial _\nu X^++\left(1-a\right)G_{\varphi i}\partial _\nu
 X^i\right]\right\},
\end{eqnarray}
where
\begin{equation}\label{}
 X^+=\left(1-a\right)t+a\varphi ,
\end{equation}
and the T-dual transverse metric is
\begin{equation}\label{}
 \tilde{G}_{ij}=\left[\left(1-a\right)^2G_{\varphi\varphi}-a^2G_{tt}
 \right]G_{ij}-\left(1-a\right)^2G_{\varphi i}G_{\varphi j}.
\end{equation}
The corresponding Nambu-Goto Lagrangian is
\begin{eqnarray}\label{ng}
 \mathcal{L}_{\rm NG}&=&
 -\frac{1}
 {\left(1-a\right)^2G_{\varphi\varphi}-a^2G_{tt}}
 \left\{\vphantom{\frac{aG_{tt}}{1-a}}
 \left[\vphantom{+\left(aG_{tt}G_{\varphi i}\partial _1X^i
 -\tilde{G}_{ij}\partial _1X^i\partial _1X^j\right)^2}
 \left(
 G_{tt}G_{\varphi \varphi }+2aG_{tt}G_{\varphi i}\partial _0X^i-
 \tilde{G}_{ij}\partial _0X^i\partial _0X^j\right)
 \right.\right. \nonumber \\
 && \left.\left.\times
 \left(1+\tilde{G}_{ij}\partial _1X^i\partial _1X^j\right)
 +\left(aG_{tt}G_{\varphi i}\partial _1X^i
 -\tilde{G}_{ij}\partial _0X^i\partial _1X^j\right)^2
 \right]^{1/2}
 \right.\nonumber \\ &&\left.
 -\frac{aG_{tt}}{1-a}-\left(1-a\right)G_{\varphi i}\partial _0X^i
 \right\}.
\end{eqnarray}
This is written in the gauge
\begin{equation}\label{}
 X^+=\tau ,\qquad \tilde{\varphi }=\sigma .
\end{equation}
The consistency of this gauge with the boundary condition (\ref{bc})
requires that the length of the string is identified with the
light-cone momentum in the units of the string tension:
\begin{equation}\label{ident}
 L=\mathcal{J}_+.
\end{equation}
Of course one can choose arbitrary $L$, but then the gauge condition
should be replaced by $\tilde{\varphi }=\mathcal{J}_+\sigma /L$, and
the factor $\mathcal{J}_+/L$ will appear in many places in the gauge
fixed Lagrangian. The choice (\ref{ident}) has an advantage that all
the coefficients in the Lagrangian are pure numbers. The only
coupling constant is an overall factor of the string tension.

The Nambu-Goto action (\ref{ng}) can be readily expanded in the
transverse fluctuations. If we assume that $G_{tt}=1+O(X^2)$,
$G_{\varphi \varphi }=1+O(X^2)$ and $G_{\varphi i}=O(X^3)$, then up
to the quartic order:
\begin{eqnarray}\label{}
 \mathcal{L}_{\rm NG}&=&
 \frac{G_{ij}}{2}\,\partial _\mu X^i\partial ^\mu X^j
 -\frac{G_{tt}}{2}+\frac{G_{\varphi \varphi }}{2}
 +G_{\varphi i}\partial _0X^i
 \nonumber \\
 &&
 +\frac{1}{4}\left(1-G_{tt}G_{\varphi \varphi }\right)
 G_{ij}\left(\partial _0X^i\partial _0X^j+\partial _1X^i\partial _1X^j\right)
 +\frac{1}{4}\left(G_{tt}-1\right)^2-\frac{1}{4}\left(G_{\varphi \varphi
 }-1\right)^2
 \nonumber \\ &&
 -\frac{1-2a}{8}\left({G_{tt}}-{G_{\varphi \varphi
 }}\right)^2
 -\frac{1-2a}{8}\left(G_{ij}\partial _\mu X^i\partial ^\mu
 X^j\right)^2
 \nonumber \\ &&
 +\frac{1-2a}{4}\left(G_{ij}\partial _\mu X^i\partial _\nu
 X^j\right)^2+\ldots .
\end{eqnarray}
The first line is the naive Lagrangian obtained by setting $t=\tau
=\varphi $ in the Polyakov action, the second line is an additional
gauge-invariant interaction that arises at the quartic order, the
last two lines are gauge-dependent and vanish in the pure light-cone
gauge $a=1/2$.

\bibliographystyle{nb}
\bibliography{refs}

\begin{thebibliography}{10}
\ifx\href\asklfhas\newcommand{\href}[2]{#2}\fi
\raggedright
\small
\parskip 0pt

\bibitem{Ahn:2008aa}
C.~Ahn and R.~I.~Nepomechie,
\textit{``{N=6 super Chern-Simons theory S-matrix and all-loop Bethe ansatz
  equations}''},
\textsf{JHEP~0809,~010~(2008)},
\href{http://arXiv.org/abs/0807.1924}{\texttt{0807.1924}}.
%
\bibitem{Aharony:2008ug}
O.~Aharony, O.~Bergman, D.~L.~Jafferis and J.~Maldacena,
\textit{``{N=6 superconformal Chern-Simons-matter theories, M2-branes and their
  gravity duals}''},
\textsf{JHEP~0810,~091~(2008)},
\href{http://arXiv.org/abs/0806.1218}{\texttt{0806.1218}}.
%
\bibitem{Aharony:2008gk}
O.~Aharony, O.~Bergman and D.~L.~Jafferis,
\textit{``{Fractional M2-branes}''},
\textsf{JHEP~0811,~043~(2008)},
\href{http://arXiv.org/abs/0807.4924}{\texttt{0807.4924}}.
%
\bibitem{Arutyunov:2008if}
G.~Arutyunov and S.~Frolov,
\textit{``{Superstrings on $AdS_4 \times CP^3$ as a Coset Sigma-model}''},
\textsf{JHEP~0809,~129~(2008)},
\href{http://arXiv.org/abs/0806.4940}{\texttt{0806.4940}}.
%
\bibitem{Stefanski:2008ik}
j.~Stefanski,~B.,
\textit{``{Green-Schwarz action for Type IIA strings on $AdS_4\times CP^3$}''},
\textsf{Nucl.~Phys.~B808,~80~(2009)},
\href{http://arXiv.org/abs/0806.4948}{\texttt{0806.4948}}.
%
\bibitem{Gomis:2008jt}
J.~Gomis, D.~Sorokin and L.~Wulff,
\textit{``{The complete $AdS_4\times CP^3$ superspace for the type IIA
  superstring and D-branes}''},
\textsf{JHEP~0903,~015~(2009)},
\href{http://arXiv.org/abs/0811.1566}{\texttt{0811.1566}}.
%
\bibitem{Minahan:2008hf}
J.~A.~Minahan and K.~Zarembo,
\textit{``{The Bethe ansatz for superconformal Chern-Simons}''},
\textsf{JHEP~0809,~040~(2008)},
\href{http://arXiv.org/abs/0806.3951}{\texttt{0806.3951}}.
%
\bibitem{Gromov:2008bz}
N.~Gromov and P.~Vieira,
\textit{``{The AdS4/CFT3 algebraic curve}''},
\textsf{JHEP~0902,~040~(2009)},
\href{http://arXiv.org/abs/0807.0437}{\texttt{0807.0437}}.
%
\bibitem{Gromov:2008qe}
N.~Gromov and P.~Vieira,
\textit{``{The all loop AdS4/CFT3 Bethe ansatz}''},
\textsf{JHEP~0901,~016~(2009)},
\href{http://arXiv.org/abs/0807.0777}{\texttt{0807.0777}}.
%
\bibitem{Gromov:2009tv}
N.~Gromov, V.~Kazakov and P.~Vieira,
\textit{``{Integrability for the Full Spectrum of Planar AdS/CFT}''},
\href{http://arXiv.org/abs/0901.3753}{\texttt{0901.3753}}.
%
\bibitem{Gaiotto:2008cg}
D.~Gaiotto, S.~Giombi and X.~Yin,
\textit{``{Spin Chains in N=6 Superconformal Chern-Simons-Matter Theory}''},
\textsf{JHEP~0904,~066~(2009)},
\href{http://arXiv.org/abs/0806.4589}{\texttt{0806.4589}}.
%
\bibitem{Bak:2008cp}
D.~Bak and S.-J.~Rey,
\textit{``{Integrable Spin Chain in Superconformal Chern-Simons Theory}''},
\textsf{JHEP~0810,~053~(2008)},
\href{http://arXiv.org/abs/0807.2063}{\texttt{0807.2063}}.
%
\bibitem{Bak:2008vd}
D.~Bak, D.~Gang and S.-J.~Rey,
\textit{``{Integrable Spin Chain of Superconformal U(M)xU(N) Chern- Simons
  Theory}''},
\textsf{JHEP~0810,~038~(2008)},
\href{http://arXiv.org/abs/0808.0170}{\texttt{0808.0170}}.
%
\bibitem{Zwiebel:2009vb}
B.~I.~Zwiebel,
\textit{``{Two-loop Integrability of Planar N=6 Superconformal Chern- Simons
  Theory}''},
\href{http://arXiv.org/abs/0901.0411}{\texttt{0901.0411}}.
%
\bibitem{Minahan:2009te}
J.~A.~Minahan, W.~Schulgin and K.~Zarembo,
\textit{``{Two loop integrability for Chern-Simons theories with N=6
  supersymmetry}''},
\textsf{JHEP~0903,~057~(2009)},
\href{http://arXiv.org/abs/0901.1142}{\texttt{0901.1142}}.
%
\bibitem{Bergman:2009zh}
O.~Bergman and S.~Hirano,
\textit{``{Anomalous radius shift in AdS(4)/CFT(3)}''},
\href{http://arXiv.org/abs/0902.1743}{\texttt{0902.1743}}.
%
\bibitem{Nishioka:2008gz}
T.~Nishioka and T.~Takayanagi,
\textit{``{On Type IIA Penrose Limit and N=6 Chern-Simons Theories}''},
\textsf{JHEP~0808,~001~(2008)},
\href{http://arXiv.org/abs/0806.3391}{\texttt{0806.3391}}.
%
\bibitem{Grignani:2008is}
G.~Grignani, T.~Harmark and M.~Orselli,
\textit{``{The SU(2) x SU(2) sector in the string dual of N=6 superconformal
  Chern-Simons theory}''},
\textsf{Nucl.~Phys.~B810,~115~(2009)},
\href{http://arXiv.org/abs/0806.4959}{\texttt{0806.4959}}.
%
\bibitem{Grignani:2008te}
G.~Grignani, T.~Harmark, M.~Orselli and G.~W.~Semenoff,
\textit{``{Finite size Giant Magnons in the string dual of N=6 superconformal
  Chern-Simons theory}''},
\textsf{JHEP~0812,~008~(2008)},
\href{http://arXiv.org/abs/0807.0205}{\texttt{0807.0205}}.
%
\bibitem{Lee:2008ui}
B.-H.~Lee, K.~L.~Panigrahi and C.~Park,
\textit{``{Spiky Strings on $AdS_4 \times {\bf CP}^3$}''},
\textsf{JHEP~0811,~066~(2008)},
\href{http://arXiv.org/abs/0807.2559}{\texttt{0807.2559}}.
%
\bibitem{Shenderovich:2008bs}
I.~Shenderovich,
\textit{``{Giant magnons in $AdS_4/CFT_3$: dispersion, quantization and
  finite-size corrections}''},
\href{http://arXiv.org/abs/0807.2861}{\texttt{0807.2861}}.
%
\bibitem{Ahn:2008hj}
C.~Ahn, P.~Bozhilov and R.~C.~Rashkov,
\textit{``{Neumann-Rosochatius integrable system for strings on $AdS_4 \times
  CP^3$}''},
\textsf{JHEP~0809,~017~(2008)},
\href{http://arXiv.org/abs/0807.3134}{\texttt{0807.3134}}.
%
\bibitem{Ryang:2008rc}
S.~Ryang,
\textit{``{Giant Magnon and Spike Solutions with Two Spins in $AdS_4\times
  CP^3$}''},
\textsf{JHEP~0811,~084~(2008)},
\href{http://arXiv.org/abs/0809.5106}{\texttt{0809.5106}}.
%
\bibitem{Bombardelli:2008qd}
D.~Bombardelli and D.~Fioravanti,
\textit{``{Finite-Size Corrections of the $\mathbb{CP}^3$ Giant Magnons: the
  L\'{u}scher terms}''},
\href{http://arXiv.org/abs/0810.0704}{\texttt{0810.0704}}.
%
\bibitem{Lukowski:2008eq}
T.~Lukowski and O.~O.~Sax,
\textit{``{Finite size giant magnons in the SU(2) x SU(2) sector of $AdS_4 x
  CP^3$}''},
\textsf{JHEP~0812,~073~(2008)},
\href{http://arXiv.org/abs/0810.1246}{\texttt{0810.1246}}.
%
\bibitem{Ahn:2008wd}
C.~Ahn and P.~Bozhilov,
\textit{``{Finite-size Effect of the Dyonic Giant Magnons in N=6 super
  Chern-Simons Theory}''},
\href{http://arXiv.org/abs/0810.2079}{\texttt{0810.2079}}.
%
\bibitem{Abbott:2008qd}
M.~C.~Abbott and I.~Aniceto,
\textit{``{Giant Magnons in $AdS_4 \times CP^3$: Embeddings, Charges and a
  Hamiltonian}''},
\href{http://arXiv.org/abs/0811.2423}{\texttt{0811.2423}}.
%
\bibitem{Kalousios:2009mp}
C.~Kalousios, M.~Spradlin and A.~Volovich,
\textit{``{Dyonic Giant Magnons on $CP^3$}''},
\href{http://arXiv.org/abs/0902.3179}{\texttt{0902.3179}}.
%
\bibitem{Suzuki:2009sc}
R.~Suzuki,
\textit{``{Giant Magnons on $CP^3$ by Dressing Method}''},
\href{http://arXiv.org/abs/0902.3368}{\texttt{0902.3368}}.
%
\bibitem{Beisert:2005tm}
N.~Beisert,
\textit{``{The su(2|2) dynamic S-matrix}''},
\textsf{Adv.~Theor.~Math.~Phys.~12,~945~(2008)},
\href{http://arXiv.org/abs/hep-th/0511082}{\texttt{hep-th/0511082}}.
%
\bibitem{Beisert:2006qh}
N.~Beisert,
\textit{``{The Analytic Bethe Ansatz for a Chain with Centrally Extended
  su(2|2) Symmetry}''},
\textsf{J.~Stat.~Mech.~0701,~P017~(2007)},
\href{http://arXiv.org/abs/nlin/0610017}{\texttt{nlin/0610017}}.
%
\bibitem{Beisert:2006ez}
N.~Beisert, B.~Eden and M.~Staudacher,
\textit{``{Transcendentality and crossing}''},
\textsf{J.~Stat.~Mech.~0701,~P021~(2007)},
\href{http://arXiv.org/abs/hep-th/0610251}{\texttt{hep-th/0610251}}.
%
\bibitem{Beisert:2006ib}
N.~Beisert, R.~Hernandez and E.~Lopez,
\textit{``A crossing-symmetric phase for $AdS_5 \times S^5$ strings''},
\textsf{JHEP~0611,~070~(2006)},
\href{http://arXiv.org/abs/hep-th/0609044}{\texttt{hep-th/0609044}}.
%
\bibitem{Ahn:2008tv}
C.~Ahn and R.~I.~Nepomechie,
\textit{``{An alternative S-matrix for N=6 Chern-Simons theory ?}''},
\textsf{JHEP~0903,~068~(2009)},
\href{http://arXiv.org/abs/0810.1915}{\texttt{0810.1915}}.
%
\bibitem{Ahn:2009zg}
C.~Ahn and R.~I.~Nepomechie,
\textit{``{Two-loop test of the N=6 Chern-Simons theory S-matrix}''},
\textsf{JHEP~0903,~144~(2009)},
\href{http://arXiv.org/abs/0901.3334}{\texttt{0901.3334}}.
%
\bibitem{McLoughlin:2008ms}
T.~McLoughlin and R.~Roiban,
\textit{``{Spinning strings at one-loop in $AdS_4 \times P^3$}''},
\textsf{JHEP~0812,~101~(2008)},
\href{http://arXiv.org/abs/0807.3965}{\texttt{0807.3965}}.
%
\bibitem{Alday:2008ut}
L.~F.~Alday, G.~Arutyunov and D.~Bykov,
\textit{``{Semiclassical Quantization of Spinning Strings in $AdS_4 \times
  CP^3$}''},
\textsf{JHEP~0811,~089~(2008)},
\href{http://arXiv.org/abs/0807.4400}{\texttt{0807.4400}}.
%
\bibitem{Krishnan:2008zs}
C.~Krishnan,
\textit{``{$AdS_4/CFT_3$ at One Loop}''},
\textsf{JHEP~0809,~092~(2008)},
\href{http://arXiv.org/abs/0807.4561}{\texttt{0807.4561}}.
%
\bibitem{Gromov:2008fy}
N.~Gromov and V.~Mikhaylov,
\textit{``{Comment on the Scaling Function in $AdS_4\times CP_3$}''},
\textsf{JHEP~0904,~083~(2009)},
\href{http://arXiv.org/abs/0807.4897}{\texttt{0807.4897}}.
%
\bibitem{McLoughlin:2008he}
T.~McLoughlin, R.~Roiban and A.~A.~Tseytlin,
\textit{``{Quantum spinning strings in $AdS_4 \times CP^3$: testing the Bethe
  Ansatz proposal}''},
\textsf{JHEP~0811,~069~(2008)},
\href{http://arXiv.org/abs/0809.4038}{\texttt{0809.4038}}.
%
\bibitem{Beisert:2005di}
N.~Beisert, V.~A.~Kazakov, K.~Sakai and K.~Zarembo,
\textit{``{Complete spectrum of long operators in N = 4 SYM at one loop}''},
\textsf{JHEP~0507,~030~(2005)},
\href{http://arXiv.org/abs/hep-th/0503200}{\texttt{hep-th/0503200}}.
%
\bibitem{Gromov:2007ky}
N.~Gromov and P.~Vieira,
\textit{``{Complete 1-loop test of AdS/CFT}''},
\textsf{JHEP~0804,~046~(2008)},
\href{http://arXiv.org/abs/0709.3487}{\texttt{0709.3487}}.
%
\bibitem{Astolfi:2008ji}
D.~Astolfi, V.~G.~M.~Puletti, G.~Grignani, T.~Harmark and M.~Orselli,
\textit{``{Finite-size corrections in the SU(2) x SU(2) sector of type IIA
  string theory on $AdS_4 x CP^3$}''},
\textsf{Nucl.~Phys.~B810,~150~(2009)},
\href{http://arXiv.org/abs/0807.1527}{\texttt{0807.1527}}.
%
\bibitem{Sundin:2008vt}
P.~Sundin,
\textit{``{The $AdS_4\times CP_3$ string and its Bethe equations in the near
  plane wave limit}''},
\textsf{JHEP~0902,~046~(2009)},
\href{http://arXiv.org/abs/0811.2775}{\texttt{0811.2775}}.
%
\bibitem{Uvarov:2008yi}
D.~V.~Uvarov,
\textit{``{$AdS_4 \times CP^3$ superstring and D=3 N=6 superconformal
  symmetry}''},
\href{http://arXiv.org/abs/0811.2813}{\texttt{0811.2813}}.
%
\bibitem{Arutyunov:2006gs}
G.~Arutyunov, S.~Frolov and M.~Zamaklar,
\textit{``{Finite-size effects from giant magnons}''},
\textsf{Nucl.~Phys.~B778,~1~(2007)},
\href{http://arXiv.org/abs/hep-th/0606126}{\texttt{hep-th/0606126}}.
%
\bibitem{Arutyunov:2009ga}
G.~Arutyunov and S.~Frolov,
\textit{``{Foundations of the $AdS_5 \times S^5$ Superstring. Part I}''},
\href{http://arXiv.org/abs/0901.4937}{\texttt{0901.4937}}.
%
\bibitem{Arutyunov:2004vx}
G.~Arutyunov, S.~Frolov and M.~Staudacher,
\textit{``Bethe ansatz for quantum strings''},
\textsf{JHEP~0410,~016~(2004)},
\href{http://arXiv.org/abs/hep-th/0406256}{\texttt{hep-th/0406256}}.
%
\bibitem{Dorey:2007xn}
N.~Dorey, D.~M.~Hofman and J.~M.~Maldacena,
\textit{``{On the singularities of the magnon S-matrix}''},
\textsf{Phys.~Rev.~D76,~025011~(2007)},
\href{http://arXiv.org/abs/hep-th/0703104}{\texttt{hep-th/0703104}}.
%
\bibitem{Klose:2006zd}
T.~Klose, T.~McLoughlin, R.~Roiban and K.~Zarembo,
\textit{``{Worldsheet scattering in $AdS_5\times S^5$}''},
\textsf{JHEP~0703,~094~(2007)},
\href{http://arXiv.org/abs/hep-th/0611169}{\texttt{hep-th/0611169}}.
%
\bibitem{Kruczenski:2004cn}
M.~Kruczenski and A.~A.~Tseytlin,
\textit{``{Semiclassical relativistic strings in $S^5$ and long coherent
  operators in N = 4 SYM theory}''},
\textsf{JHEP~0409,~038~(2004)},
\href{http://arXiv.org/abs/hep-th/0406189}{\texttt{hep-th/0406189}}.
%
\end{thebibliography}

\end{document}